# INFORMATIONAL EXPERIMENTS WITH MICROPARTICLES AND ATOMS[1]

**Raoul Nakhmanson[2]**

*Institute for Didactics of Physics*
*Justus-Liebig University*
*6300 Giessen, Germany*

Accepting information as a physical category and ascribing to inanimate matter some spirit (consciousness, intelligence) allows to explain quantum-mechanical phenomena, including delayed-choice and EPR-Bohm-Bell experiments, as well as irreversibility of time, remaining on the basis of local realism, and suggest essentially new experiments with microparticles and atoms in which information plays the principal role.

Historically formed physics is only a part of human knowledge. It searches laws of behaviour of so-called "inanimate" matter, and does it with a success. The behaviour of "animate" matter is also a subject of physics, but only if it does not include spiritual processes. The consequent departments of physics are biomechanics and biophysics. The spiritual phenomena lay beyond the limits of the traditional physics. In this respect many prominent physicists, including Einstein and Bohr, had similar opinions.

The division of *"being"* in spirit and matter is ascribed to Descartes. The sages of ancient times (and of the modern East) did not make such a division and thought all matter was animated. For example, Aristoteles thought that any matter had not only *"potentia"* - ability to move, but also *"entelechia"*—a wish and purpose to move. I think this is a correct point of view. Spirit can not be separated from matter, a boundary line between them cannot be drawn. The Cartesian dualism is only a very useful method of analysis, no more.

Omitting the middle age and the golden age of classical physics, I jump to the transition period and cite Emil Borel, an acquaintance of the de Broglie family. In the book *Le Hasard*, isued in 1913, the year of Bohr's atom model, Borel remarked that humans cause entropy to decrease in small volumes by means of processes accompanied by increasing of entropy in big volumes, and then goes on as follows:

> In other words, the structure of the universe is becoming more and more subtle..., it is probable that similar phenomena are going on on other scales as well, too large or too small to be accessible to us. Thus, the evolution of the universe may be represented as a gradual complication of its structure, accessible to understanding and use of beings of lesser and lesser size. As there is no absolute standard of length, we may not be afraid of such a lessening of scales; it seems to us presently that beings of molecular sizes and, all the more, beings so small in respect to molecules as we are to sun, are objects scarcely deserving our attention; but it is quite possible that the progressing complexity of the universe will create or has already created some beings with an organisation much more complex than ours.

Ascribing to "inanimate" matter some spirit (other terms: consciousness, intelligence) is not alien to modern science. Similar meanings were expressed by well known physicists, such as Charles Galton Darwin, Eddington, Heisenberg, Schrodinger, Jordan, Margenau, Wigner, Cochran, and others. [1]

Quantum-mechanical indeterminism looks like free will, i.e., it allows a micro-object to choose freely among several possible alternatives. But quantum mechanics does not search this way, it only postulates indeterminism, and *a priori* declares it as absolutely random. Thus, quantum mechanics closed its eyes on possible purposefullness of any objects behaviour, *a priori* replacing it by some chaos.

The notion "information" is a good connecting link between physics and the humanities. Firstly, it is used in communication technique and is formalized in the "theory of information." Secondly, it calls forth some suppositions about sources and receivers of information which are intelligent or supported by some intelligence. Thirdly, interchange and treatment of information are the essence of spiritual life of individuals and society as a whole.

Is information of physical category? Or, in other words, does information play a role in a world of "inanimate" objects? This question, in its latent form, has existed in physics for more than hundred years and is connected with discussions around Maxwell's demon of thermodynamics, the passive observer of the theory of relativity, the active experimenter of quantum mechanics, and the coincidence of Boltzmann's and Shannon's formulae for entropy. Accepting information as an inherent attribute of matter, we ascribe to "inanimate" matter some intelligence. This question has a direct relation to irreversibility of time and to the crisis of notions of reality and locality in modern physics.

The clues supporting the opinion that information like mass and energy is an inherent attribute of matter (which, in it's turn, is intelligent) are the following:

(1) the Heaviside-Einstein formula $E = mc^2$, because $c$ is the maximal velocity for the spreading of information;
(2) the informational character of the wave function $\Psi$, describing probability, in contrast to classical potentials describing "tangible" fields transporting energy and impulse;
(3) "teleological" movement of matter seemingly in the principle of the least action; and
(4) quantum-mechanical stochastics which can be seen as optimal tactics of behaviour of disconnected members of a quantum-mechanical ensemble searching all possible alternatives.

The irreversibility of time and the second law of thermodynamics can be explained if the state with maximal entropy has some purposeful advantages. For example, molecules of gas "inspect" the space better when they are distributed in a vessel homogeneously. Accordingly, they keep such a distribution (using control of collision parameters) and do not gather.

The acceptance of spirit or even intelligence of microparticles leads us to the acceptance of their complex structure. Do they have enough capacitance? Theoretically it was shown (firstly perhaps by Markov [2]) that one microparticle can contain a whole universe ("fridmon," "baby-universe"). If one goes not farther than Planck length ($\approx 10^{-35}$ m), one finds in a micro-object ($\varnothing \approx 10^{-15}$ to $10^{-20}$ m) of the order of $10^{45} - 10^{60}$ "Planck cells," that is much more not only than the number of neutrons in the human brain but than the number of atoms contained in all living beings, too.

A consistent development of this idea permits explaining quantum-mechanical phenomena remaining on the basis of local realism, and suggests essentially new experiments with microparticles and atoms in which information plays the principal role.

The $\Psi$ function in this conception is a strategy function. It reflects an optimal behavior of particles, both individual and social (ensemble, universum). Where is this function? Of course, it is not in the real 3-dimensional space; it is in imaginary configurational space, which, in its turn, is in the imagination (consciousness) of the particle. When the particle receives new information (it takes place by any interaction with micro- or macroobjects), it changes its strategy. Thus occurs the so-called "collapse" of the $\Psi$-function. It occurs in the consciousness of the particle (but not in the observer's consciousness, as was suggested by von Neumann and Wigner), i.e., locally and realistically. In a human consciousness collapses only its knowledge about the $\Psi$-function.

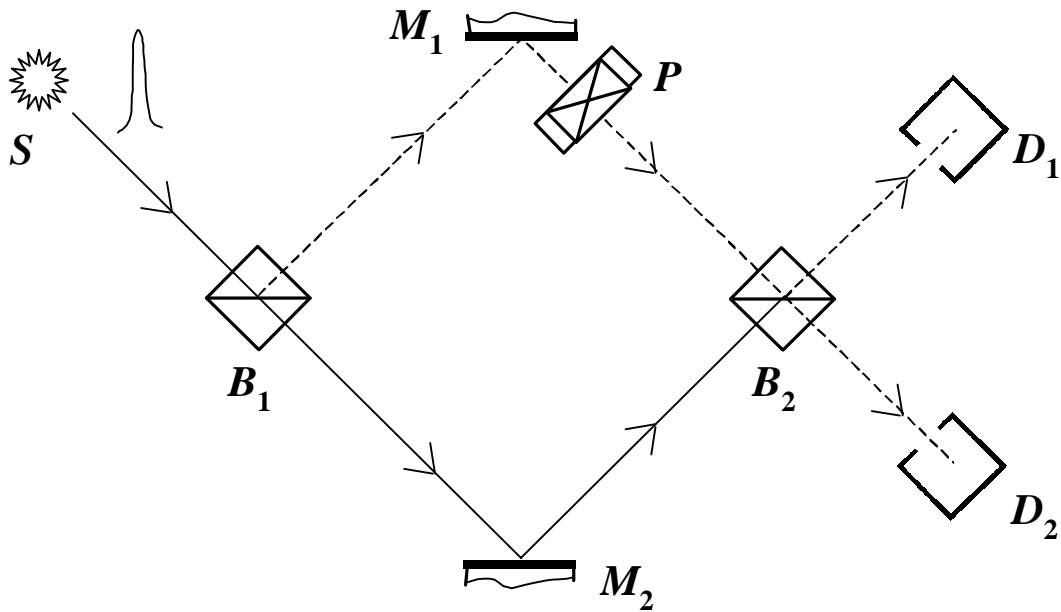

Fig. 1. "Delayed-choice" experiment with Mach-Zender interferometer. $S$ is the source of photons; $B_1$, $B_2$ are the beam-splitters; $M_1$, $M_2$ are the mirrors; $D_1$, $D_2$ are the detectors; $P$ is the Pockels cell.

The flows of particles carry not only mass and energy, but also information. It must be emphasized that the Shannon formula for information gives only a "static" approximation. In accordance with this formula, today's and yesterday's newspapers contain approximately the same information. The complex object having the same mass, energy, entropy, and Shannon information can contain different "actual" information. The different states of such an object as a functional series of time are an analogue of the series of newspapers. Therefore it is possible during the interaction to transfer information without a change of the mass, energy, and entropy of interacting objects. These flows of "actual" information mean the "spirit" life of matter.

Properly speaking, the same is true for traditional live ojbects, e.g., human beings. The flows of mass, energy and entropy have essential convergences only for growing and learning children. For an adult and educated person these flows pass mainly through him and change only the "actual" information.

Let me explain this by the example of the "delayed choice" experiment with the Mach-Zender interferometer, Fig. 1, which was discussed by Wheeler [3] and really performed in Maryland, Munich, and Frankfurt [4,5]. Here $S$ is the source of particles (photons), $B_1$ and $B_2$ are the beamsplitters, $M_1$ and $M_2$ are the mirrors, $D_1$ and $D_2$ are detectors, and $P$ is the Pockels cell. The flow of photons is so weak that no more than one photon is in the interferometer as a rule.

The Pockels cell is at all times open (closed), but anytime for a short moment (several nanoseconds) when a photon can pass it, it is closed (opened). As a result of experiments having been performed, the interference after $B_2$ appears only if the Pockels cell was open at the appropriate moment, independent of its state at other times.

This result can not be explained by remaining on the level of classical realism and thinking that, after $B_1$, a photon really chooses one path (upper or lower in Fig. 1) in the interferometer, because if the photon really traveled the lower path (this "must be" in 50% of the cases), it is too far (several meters) from the Pockels cell, and at the above mentioned "appropriate moment" is under no influence from the cell (superluminal action is supposed to be impossible). The photon in the interferometer looks like a "Great Smoky Dragon" (Fig. 2, drawing by Field Gilbert).

From a new point of view all this looks different. When a photon meets the first beamsplitter $B_1$, it has a choice. Simultaneously it receives from $B_1$ the fresh information about the *past* of the world, particularly about the interferometer and Pockels cell. The physical conditions of $B_1$ induce a 50% choice. But the decision must be at random: It is the optimal tactics for an ensemble of disconnected photons. The consciousness of the photon works only to find an optimal strategy, i.e., $\Psi$ function (in this case a 50% choice of path after $B_1$, and interference after $B_2$, if Pockels cell appears open). To find it, the photon solves a variation problem (e.g., the wave equation).

Let us suggest that after $B_1$ the photon takes the lower path. It meets the mirror $M_2$ and becomes some new information about the *past* of the world. But in respect to interferometer, it is the same information, so this part of the wave function stays the same. Finally our photon meets the beamsplitter $B_2$ and must again choose its path. Simultaneously it receives from $B_2$ the fresher information about the *past* of the world, including an actual history of the upper arm of the interferometer with the state of Pockels cell at the "appropriate moment." This information is brought to $B_2$ with luminar velocity, e.g., by virtual photons in solids and by thermal photons in vacuum/air: At room temperature the characteristic time of several nanoseconds is enough for $B_2$ to receive from $P$ several thousands of thermal photons. Actually, the number of thermal photons emitted by one and arriving at other macroobject is

$$N \approx 5 \cdot 10^{10} \, S_1 \, S_2 \, T^3 \, \tau / L^2,$$

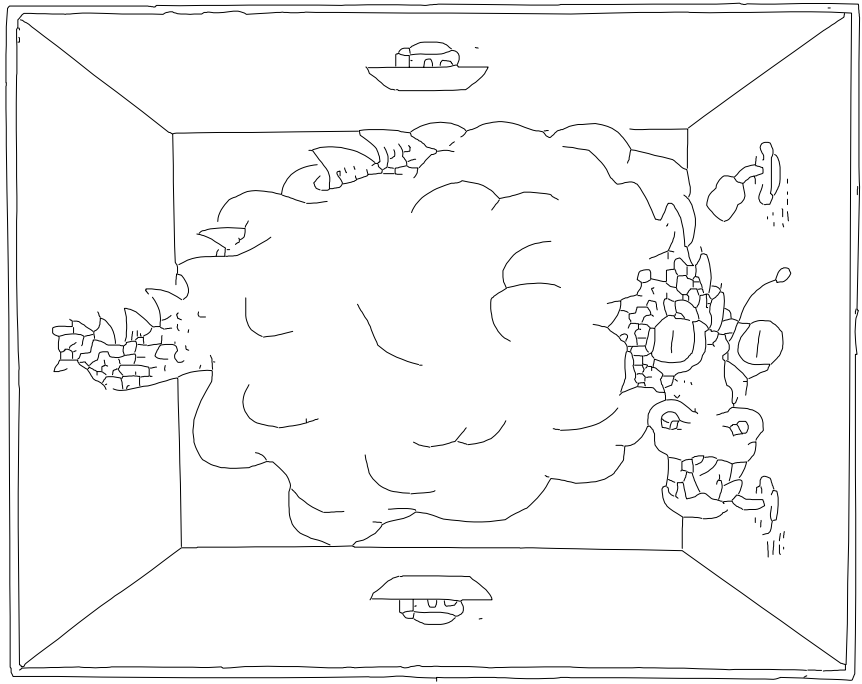

Fig. 2. "Great Smoky Dragon" (from Ref. 4; drawing by Field Gilbert). Within the standard (Copenhagen) interpretation, in spite of our knowing of input and output of a photon (the tail and the mouth of the dragon, respectively), we neither know nor have the right to say how the photon came. As it is explained in the text, this dragon exists only in the photon's imagination (and in imagination of physicist), not in the interferometer.

where $S_1$ and $S_2$ are the effective areas of the first and the second macro-objects, respectively, $T$ is the absolute temperature of $S_1$, $\tau$ is the characteristic time, and $L$ is the distance between $S_1$ and $S_2$. Substituting the typical values $S_1 = S_2 = 2$ cm$^2$, $T = 293°$K, $\tau = 10^{-8}$ sec, and $L = 10^3$ cm, one finds $N \approx 5000$.

Now our photon has all the necessary information to make a decision, namely, to prefer a direction of constructive interference (if the cell at the "appropriate moment" was opened) or to make a 50% random choice between two possible directions (if the cell at the "appropriate moment" was closed). If the state of the Pockels cell in the "appropriate moment" was different than before, a reduction of the strategy (i.e., a reduction of wave function $\Psi$ relating to the experiment) takes place in the consciousness of the photon after interaction with the beamsplitter $B_2$. The "Great Smoky Dragon" does not exist in the real interferometer. One can speak only about the Dragon's images in the consciousness of the particle (primarily) and that of the physicist (secondarily).

To have a success in the "delayed choice" experiments, one can try to cut off the informational contact between the Pockels cell and the beamsplitter $B_2$ e.g. by introducing a deep-cooled filter. If the filter is cooled by liquid helium, no photons come from it to $B_2$ during the characteristic time (several nanoseconds). Besides, to prevent a prediction of a state of Pockels cell by the $B_2$, it is better to control the cell using not a regular but "good" random generator.

The situation discussed by Einstein, Podolsky, and Rosen, and in modern form by Bohm and Bell, is more complex. Here two particles flying in the opposite directions have a common correlated wave function = common correlated strategy. As a consequence, the result of interaction of one particle with some measurement apparatus must be correlated with the result of the interaction of another particle with another measurement apparatus, in spite of a large space separating the two apparatuses. Such correlation is possible only if each apparatus knows the state of the other one; more correctly, if each apparatus (and/or the particle interacting with it) can *predict* the state of the other apparatus at the "appropriate moment" of measuring with a good probability. In the world of communicating matter, like in human and other biological societies, such prediction is a natural attribute of existence, as well as the common wave function reflecting this intercommunication of whole matter of the universe. In all experiments made up to now, such *predictions* could take place, without supposing a super-luminal velocity of communication, because the states of the measurement apparatuses have been changed very slowly [6] or periodically [7]. Again one can try to perform such experiments with a "good" random control of measurement apparatuses.

It is interesting to note that the authors of experimental researches felt the advantage of random control (as it seems more intuitive, because they do not discuss it) and sometimes used it [4]. The peculiarity of random signal series is non-predicting of its next term. Therefore, these authors felt a possibility of "inanimated" matter to predict the future, and have tried to restrict it.

In the "delayed choice" and EPR experiments one tries to eliminate the informational contacts already existing in nature. It is possible to go further and try to "speak" with a quantum object itself. The first suggestions of such experiments have been made in [8]. In this case a content and code of a "message" are defined by the experimenter, making his task more difficult but more interesting. One can try, e.g., to use some universal language, like mathematics or music, and hope for an "attractivity" of information, even if it can not be decoded. Using feedback, the sending of a new message can be made dependent on the "free-choice" of a microparticle when it passes through the beamsplitter or jumps from one energy level to others.

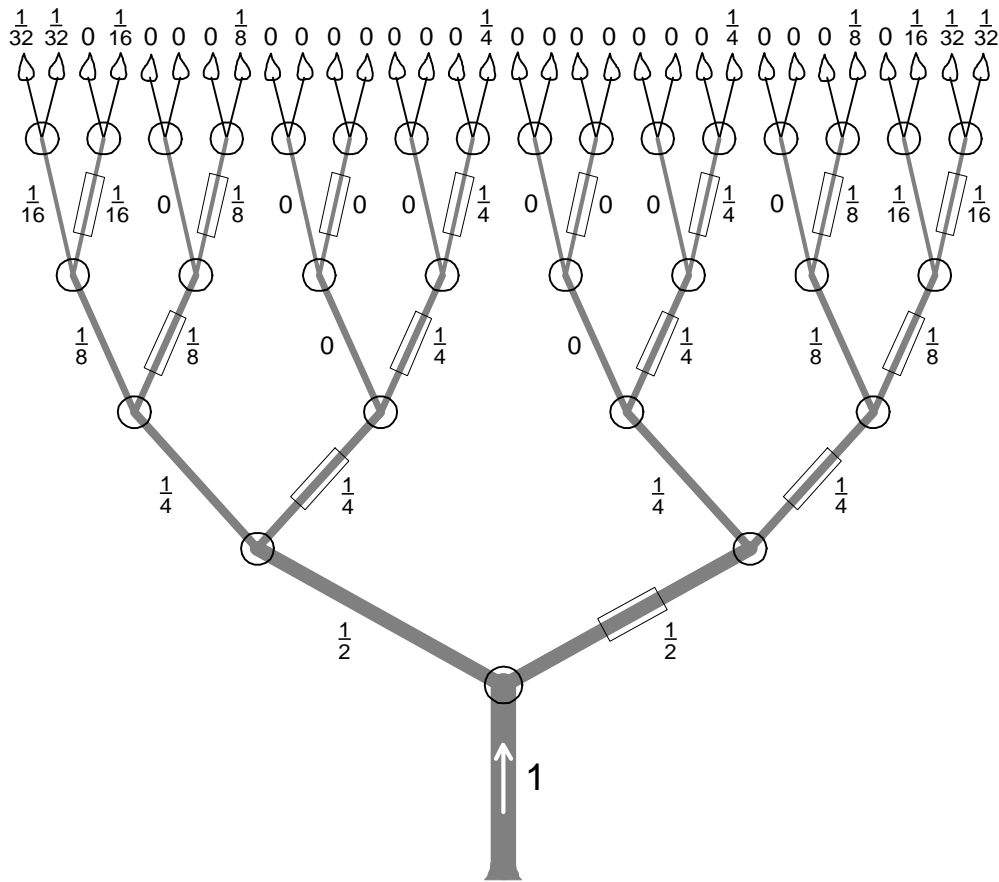

Fig. 3. Binary-tree experiment. Circles stand for beamsplitters, rectangles denote informational cells. Ciphers show the probability of detecting a particle in the case of the most rapid formation of a rigid, conservative condititional reflex.

Figure 3 shows the scheme of "binary-tree" experiment. The initial beam of micro-objects (particles, atoms) enters into a system of beam-splitters (shown by circles). They can be semi-transparent/semi-reflecting mirrors for photons, crystals for electrons or neutrons, Stern-Gerlach apparatuses for atoms, etc. Figure 3 shows only five rows of beamsplitters, but, in principle, there can be as many as experimentally feasible. According to present-day theoretical ideas and practical experience, each of the output beams has the same intensity, namely, 1/32 of intensity of the initial beam (real beamsplitters may have, of course, some absorption, but here it is not a matter of principle).

Now let us introduce into each of the right channels an "information cell" (shown by rectangles), which is a device leaving unchanged the intensity of the beam passing it, but offering some information to particles. For polarized photons such a cell may be a set of transparent plates fixed at the Brewster angle, and the information can be coded, say, by differences in the materials of plates, their thickness, or distance between them. For example, the information cell may include 10 glasses of thickness *A* and 10 glasses of thickness *B* arranged in the following way:

*BBABBBABBA   AABAAABAAB.*

In binary code (direct and inverse) they reprsent the number 0010001001 = 137, which is the reciprocal of the fine-structure constant. Absorption brought in by the information cells may be compensated by introducing into each of the left channels a compensating cell bearing no information, more precisely, bearing less, or less significant - in our judgment - information (e.g., the same glasses staying random or periodically). The information in each next row is a sequel of that in the preciding row.

   The commonly accepted point of view is that the introduction of information cells, together with compensators, will not change the uniform probability distribution of particles in the output beams. But if particles are intelligent, and are able to notice the information offered to them, they may become interested in it. After a number of rows, the particles should notice that the information is offered only in right channels, and should prefer the choice of right channels in passing through the following beamsplitters. In other words, particles could develop a "conditional reflex," of essentially the same kind as in behaviour experiments on living beings. Such an intelligent behaviour of particles should lead to a change of their distribution in the output beams. For example, if the conditional reflex appears immediately and the particles are "conservative", i.e., they are no more of interest to the left channels, the distribution of probability in branches of the binary tree is like the one shown in Fig. 3.

   Deviation from the uniform distribution of particles in the output beams will mean that the particles at least recognize the offered information and have an interest in it. This, however, still does not mean that the particles understand the information offered: People of the modern era were interested in ancient hieroglyphic symbols long before they learned how to interpret them. To establish a deciphering stage, one can, starting from some row of the binary tree, introduce some specific "requests" into the information cells. For example, one can "ask" particles to choose a left channel after the next beamsplitter rather than a right channel. The honouring of such kind of requests can easily be detected by an experimenter. However, the possibilities of an experiment typified in Fig. 3 are not exhausted by this passive level of communication. Purposefully choosing direction at each subsequent beamsplitter, the particle, in its turn, can send information to the experimenter using "right" and "left" as a binary code.

   The experiment Fig. 3 can be called "coordinate-impulse" one to distinguish it from the "energy-time" experiment which scheme is shown in Fig. 4. Here a four-level quantum system, e.g., an atom, with one low (1), one high (2), and two intermediate (3,4) energy levels is pumped by intensive radiation inducing the 1→2 transition, so that the atom stays not in the state 1 but immediately goes into the state 2. From it, the atom transits (spontaneously or light induced) to the state 3 or 4, and later transits to the state 1 completing the cycle. The radiations corresponding to some of transitions 2→3, 2→4, 3→1 and 4→1 are detected (in Fig. 4 two detectors are shown). Besides, there is an informational action on the atom, e.g., by modulation of light coming from the source S. The modulator M is controlled by the source of information SI, which, in turn, is connected with one or more detectors to close the feedback loop.

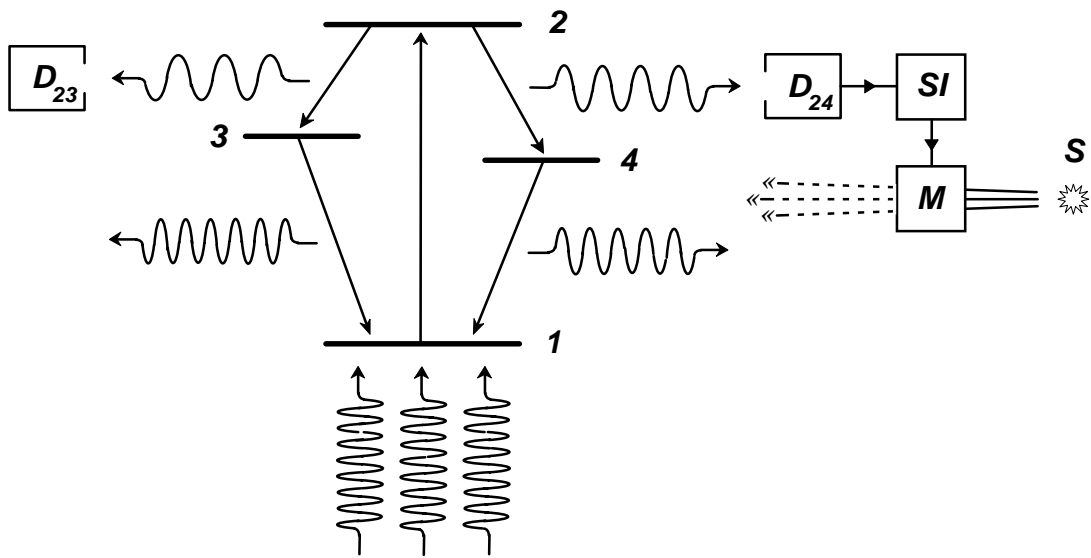

Fig. 4. Informational experiment with a single atom. 1, 2, 3, 4 are the energy levels; $D_{23}$, $D_{24}$ are detectors; S is the source of light; M is the modulator; SI is the source of information.

The feedback works in such a way as to stimulate a rate and channel of transitions, in the case of Fig. 4, the 2→4→1 transitions. The source SI sends repeatedly the same message, for example, one line of a page or one circle of gramophone record/compact disc. Only when SI receives a signal from detector $D_{24}$ it changes the message to a new one, namely, the next line or the next circle. If the atom has an intelligence and is interested in new information, it develops a conditional reflex and will prefer the 2→4 transition to the 2→3 one. Besides, the rates of both 2→4 and 4→1 transitions must increase. All this can be registered by the experimenter. To be sure that the effect is connected with information, one can make a control experiment replacing information by "indefinite" noise, etc.

If the detector $D_{24}$ has a small aperture, the space orientation of emitted photons toward to $D_{24}$ can also develop itself as a part of conditional reflex. In spite of this interesting possibility, one must prefer to use effective detecting of emitted photons to facilitate the developing of conditional reflex. Perhaps the combination of an ion trap (e.g., Penning or Paul trap) and resonator ("single atom MASER" [9]) gives a good condition for informational experiments with single atoms.

It seems that a progress in semiconductor device technology can also be used for the same experiments. In a small metal-oxide-silicon field-effect transistor (MOSFET) one can observe random telegraph signals corresponding to charge and discharge of one electron trap locating at the Si-$SiO_2$ interface [10]. As in the previous case, the new information may be sent to MOSFET (i.e., to the trap), by modulation of light or RF, only after the next capture or/and emission of an electron by the trap. It closes the feedback loop, and the experimenter may find increasing of the capture or/and emission rate.

Like the scheme of Fig. 3, in two last cases one may hope to observe not only an interest of a quantum object (atom, "surface state") to receive a new information, but decoding of it also, as well as sending of messages from the objects to the experimenter.

The noted physician Erasmus Darwin (grandfather of the great naturalist Charles Robert Darwin) played the flute for his flowers. He did not see any reaction, but after two hundred years his idea was vindicated experimentally. As it seems, this example encourages us to respect of intuition of physicist Charles Galton Darwin (Ch. R. Darwin's grandson), who wrote in 1919:

> The great positive successes of the quantum theory have accentuated all along, not merely its value, but also the essential contradictions over which it rests.... It may be that it will prove necessary to make fundamental changes in our ideas .. .or even in the last resort to endow electrons with free will.

Namely we should play music for electrons. Why not?

**NOTES**

1. Shortened version of a report which was read on September 29, 1992 in Trani Workshop (Italy).
2. Present address: Waldschmidtstrasse 131, 60314 Frankfurt, Germany.